\documentclass[amsmath,amssymb,nofootinbib,superscriptaddress]{revtex4}

\usepackage{graphicx}

\begin{document}


\vspace*{1.5cm}

\title{Hawking radiation as irreversible tunneling through self-interaction}

\author{Wen-Yu Wen}\thanks{%
E-mail: steve.wen@gmail.com}
\affiliation{Department of Physics and Center for High Energy Physics, Chung Yuan Christian University, Chung Li City, Taiwan}
\affiliation{Leung Center for Cosmology and Particle Astrophysics\\
National Taiwan University, Taipei 106, Taiwan}

\begin{abstract}
In this letter, we study the self-force in the Parikh-Wilczek tunneling model of Hawking radiation for Reissner-Nordstr\"{o}m black holes.  We conclude that the repulsive self-force speeds up the emission and the radiation becomes an irreversible process.  We also find an upper bound of charge-mass ratio for emitted particles.
\end{abstract}


\maketitle

\section{Introduction}

It was suggested by Parikh and Wilczek\cite{Parikh:1999mf} to regard Hawking radiation as a tunneling process.  The proces has its classical similarity to the spontaneous emission from a charged conductor.  The backreaction (self-gravitation) is included in the semiclassical computation and the radiation is only thermal at the leading term in expansion with respect to emission energy.  In particular, the same treatment was also applied to the radiation of a charged particle in the Reissner-Nordstr\"{o}m black hole\cite{Zhang:2005xt}\footnote{An early study of radiation of a charged particle in the Reissner-Nordstr\"{o}m black hole was given in \cite{Kraus:1994fj}.  However, the word "self-interaction" adopted there simply refers to the Coulumb interaction between the radiated charged particle and black hole from which it radiates.  The self-interaction or self-force adopted in this paper means the non-Coulumb electromagnetic interaction in curved spacetime as shown in equation (\ref{self_force}). }.  In all situations, the tunneling process respects the energy conservation law and is a reversible process\cite{Parikh:2004rh,Zhang:2009jn}.  

Recently, an universal charge-mass bound was proposed by Hod in the process of absoprtion of a charged test particle in the RN black hole\cite{Hod:2010zk}.  The essential element in his derivation is the inclusion of a term reagrding self-interaction of a charged particle in the circumference of a charged black hole.  It is curious for us whether such a bound can also be observed in the tunneling process with self-interaction.

In this letter, we study the self-force experienced by the emitted charged particles from a Reissner-Norstr\"{o}m black hole.  We find that the tunneling process will speed up and become microscopically irreversible by including the effect of self-interaction.  We also reveil similar mass-charge ratio bound by requiring unitarity.


\section{The origin of self-interaction}
The influence of gravitational force on a charged test particle at the horizon has been investigated over last decades\cite{Linet:1976sq,Smith:1980tv,Lohiya:1982fp,Bekenstein:1999cf} and can be summarized in a repulsive force:
\begin{equation}\label{self_force}
F_{self} = \frac{G_NMq^2}{c^2 r_+^3},
\end{equation}
where $M$ and $r_+$ are the black hole mass and horizon; $q$ is the charge of test particle.  This force can be attributed 
to an image charge of the same kind induced behind the black hole horizon.  We remark that this self-force exists regardless whether the black hole is charged or not.  However, in order to have radiated particles charged, we ask for a Reissner-Nordstr\"{o}m black hole.  In th following, we will adopt the units such that Newton constant $G_N$, Coulomb constant $k_Q$ and speed of light $c$ are all unities.  That this force vanishes as $M\to 0$ implies that it is the gravitational field that modifies the long-range Coulomb force of a charged particle in such a way that a finite self-force is experienced.  The additional work done by self-interaction (\ref{self_force}) for an infinetesimal change of black hole mass and test charge can be computed as:
\begin{equation}
dW = \frac{q^2}{2r_+^2}dM + \frac{Mq}{r_+^2}dq.
\end{equation}

\section{Tunneling model of charged black holes with self-interaction}
The tunnelling model of Reissner-Nordstr\"{o}m black hole radiation was first studied in \cite{Parikh:1999mf} for neutral particles and later it was generalized to the case for charged particles in \cite{Zhang:2005xt}.  Following their notation, the emitted particle with mass $\omega$ and charge $q$ follows the trajectory
\begin{equation}\label{eqn:trac}
\dot{r} = \frac{1}{2r}\frac{r^2-2(M-\omega)r+(Q-q)^2}{\sqrt{2(M-\omega)r-(Q-q)^2}},
\end{equation}
where the back reaction has been considered.  To obtain the tunneling rate, we first compute the imaginary part of action by the WKB approximation:
\begin{equation}
\text{Im S} = \text{Im} \int_{r_i}^{r_o} \int_{(M,Q)}^{(M-\omega,Q-q)}\big( \frac{dH_r}{\dot{r}}-\frac{dH_{A_t}}{\dot{r}}\big ) dr,
\end{equation}
where $r_i=M+\sqrt{M^2-Q^2}$ and $r_o=(M-\omega)+\sqrt{(M-\omega)^2-(Q-q)^2}$ are the (outer) horizon before and after emission.  The energy change due to emission of infinitesimal mass and charge are given by $dH_r$ and $dH_{A_t}$ respectively:  
\begin{eqnarray}
&&dH_r = (1 + \frac{{q^\prime}^2}{2r^2}) (-d\omega^\prime),				\nonumber\\
&&dH_{A_t} =  (\frac{Q-q^\prime}{r} + \frac{(M-\omega^\prime)q^\prime}{r^2}) (-dq^\prime).
\end{eqnarray}
We remark that the second terms in each line is the energy loss due to self-interaction.  After carrying out the contour integral with respect to $r$ and then line integral in configuration space $(\omega^\prime,q^\prime)$ , one obtains\footnote{There was some subtilty regarding this contour integral for it is only a one-way integral\cite{Akhmedov:2006pg}.  However, this subtilty was later fixed by same group of authors by considering temporal constribution\cite{Akhmedova:2008dz}.} 
\begin{eqnarray}
&& \text{Im S} = -\frac{\Delta S_{BH}(\omega,q)}{2}-F_{\gamma}(\omega,q),\nonumber\\
&& F_{\gamma}(\omega,q) \equiv \pi\int_{(0,0)}^{(\omega,q)} \frac{-{q^{\prime}}^2d\omega^\prime+2(M-\omega^\prime)q^\prime dq^\prime}{2\sqrt{(M-\omega^\prime)^2-(Q-q^\prime)}}.
\end{eqnarray}
Here $\Delta S_{BH}(\omega,q)=\pi (r_o^2-r_i^2)$ is the change of Bekenstein-Hawking entropy after radiation of a charged particle.  To evaluate the line integral $F_\gamma(\omega,q)$, a specific path $\gamma$ connecting $(\omega^\prime,q^\prime)=(0,0)$ to $(\omega,q)$ is chosen, which reflects the fact that the self-force is nonconservative.  The tunneling rate is then given by 
\begin{equation}
\Gamma \sim \exp(-2 \text{Im S}) = \exp(2F_{\gamma}(\omega,q))  \exp(\Delta S_{BH}(\omega,q)).
\end{equation}

\begin{figure}
\includegraphics[width=0.45\textwidth]{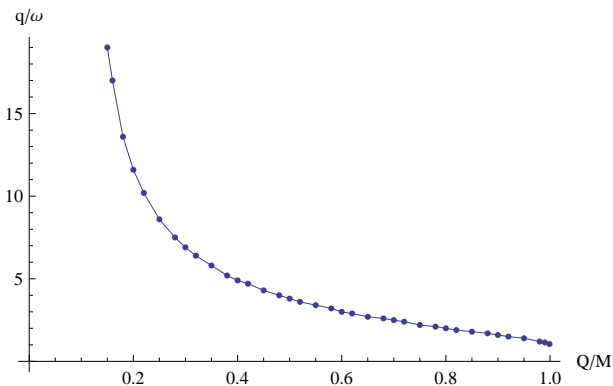} 
\caption{Plot of $|q|/ \omega$ ratio bound against various $|Q|/M$ for charged black holes.  Here we use $\gamma=1$ and $M_\omega=100$.  Here we display bounds for  $0.15 <|Q|/M < 1$.}\label{fig:fig1}
\end{figure}

Several remarks are in order:
\begin{itemize}
\item The emitted particle must carry charge in order to experience self-force, that is  $F_\gamma(\omega,0)=0$ for a neutral particle.

\item We will demonstrate that $F_\gamma(\omega,q)$ is positive for a family of integration paths in the appendix.  The positive $F_\gamma(\omega,q)$ will enhance the tunneling rate and therefore speed up the radiation.  This is consistent with the repulsive nature of self-force.

\item The total entropy change with self-interaction, $\Delta S = \Delta S_{BH} + 2 F_\gamma(\omega,q)$, is greater than $\Delta S_{BH}$.  This implies the radiation is an irreversible process even microscopically.

\item The unitarity condition states that the tunneling rate $|\Gamma| \le 1$.  This imposes an upper bound for $F_\gamma(\omega,q) \le -\frac{\Delta_{BH}}{2}$.  Suprisingly, this in turn sets up an upper bound for the ratio $q/\omega$ for emitted particles, given particular black hole mass and charge.  We plot our numerical results in the figure \ref{fig:fig1}.  The $|q|/ \omega$ ratio is bounded from above for $0.15<|Q|/M < 1$ according to our numerical results.  The bound is believed to persist for smaller $|Q|/M$ ratio and becomes unbounded for Schwarschild black holes, if the computation limit were overcome.  The bound is also found to approach unity at the extremal limit, which agrees with that derived from mutual information\cite{tba}.

\item The emitted particle during tunneling speeds up according to (\ref{eqn:trac}).  The static self-force (\ref{self_force}) may need a correction of oder ${\cal O}(1/r^4)$ for its acceleration.  We simply ignore this subleading effect here.

\end{itemize}

\appendix
\section{Positivity of $F_\gamma(\omega,q)$}
\begin{figure}
\includegraphics[width=0.45\textwidth]{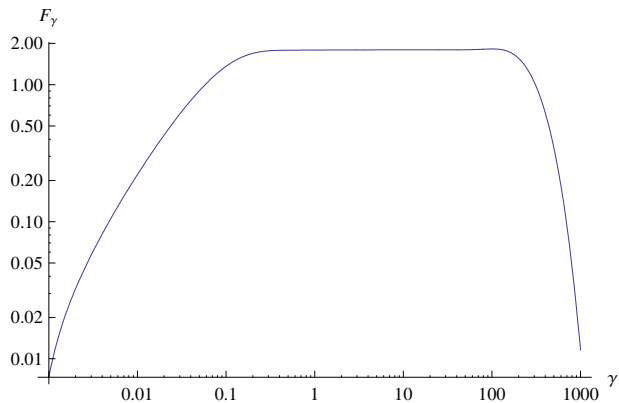} 
\caption{$F_\gamma$ is positive for $\gamma$ of all ranges and is insensitive to the choice of path when $0.5 \lesssim \gamma \lesssim 100$.}\label{fig:fig2}
\end{figure}

\begin{figure}
\includegraphics[width=0.45\textwidth]{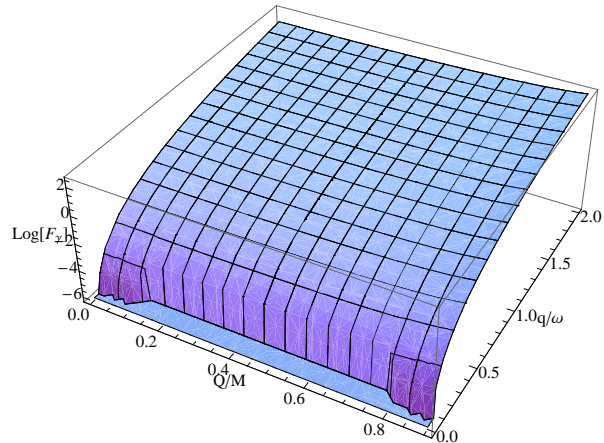} 
\caption{Plot of $F_{\gamma}$ in Log scale for various $|q|/\omega$ and $|Q|/M$ ratios.  Here we use $\gamma=1$ and $M_\omega=100$. }\label{fig:fig3}
\end{figure}

In this section we would like to show the positivity of integral $F_\gamma(\omega,q)$ over a family of paths from $(0,0)$ to $(\omega,q)$.   Before we do the explicit calculation, there are several reasons to make sense of positive $F_\gamma(\omega,q)$.  If it were negative, then first, the emission rate were slowed down for some {\sl attractive} force, and this would not caused by the desired repulsive self-force.  Secondly, the released information (total entropy loss) carried by emitted particles were smaller than the change of black hole entropy, and one would have to come up with an explanation on  that missing information.  If the $F_\gamma(\omega,q)$ happened to be zero for some path(s) $\gamma$, one could have the fine-tuning problem: why this particular path(s) could magically recover a conservative potential from a nonconservative force?  Now we seek to the numerical proof of its positivity.  We consider a class of smooth paths parametrized by $\gamma$:
\begin{equation}
\frac{q^\prime}{q} = (\frac{\omega^\prime}{\omega})^{\gamma}, \qquad 0 < \gamma < \infty.
\end{equation}
Then the integral can be expressed in terms of dimensionless variables:
\begin{eqnarray}
&&\omega^{-2}F_\gamma(\omega,q)= \pi r^2 \int_0^1\frac{[\gamma M_\omega-(\gamma+1/2)x]x^{2\gamma-1}}{\sqrt{(M_\omega-x)^2-(Q_\omega-rx^{-\gamma})^2}}dx,\nonumber\\
&&M_\omega \equiv \frac{M}{\omega}, \quad Q_\omega \equiv \frac{Q}{\omega}, \quad r \equiv \frac{q}{\omega}.
\end{eqnarray}
We carry out the integral for various ratios $|Q|/M$ for black holes and ratios $|q|/\omega$ for emitted particle, as well as for all values of admitted $\gamma$.  As shown in the figure \ref{fig:fig2}, it is found that $F_\gamma$ is always positive and insensitive to the paths for $0.5 \lesssim \gamma \lesssim 100$.  In particular, we plot $F_\gamma$ for the choice $\gamma=1$ in the figure \ref{fig:fig3}.  Since the physical mass and charge of an emitted particle is $\omega$ and q, the intermediate states $(\omega^\prime, q^\prime)$ along each integral path are regarded as virtual.  It might be possible to adopt the path integral formalism to sum over all possible configurations.  This is beyond the scope of this letter and left for future inspection.

\begin{acknowledgments}
The author is grateful to useful discussion with Hirosi Ooguri and Wei-Ming Chen in the early stage. This work is supported in part by the Taiwan's National Science Council under grant number NSC 102-2112-M-033-003-MY4 and the National Center for Theoretical Sciences.
\end{acknowledgments}


\bibliography{apssamp}

\end{document}